\documentclass[onecolumn,nofootinbib,showpacs,aps,10pt]{revtex4}

\usepackage{amssymb}
\usepackage{amsmath}
\usepackage[dvips]{color}
\usepackage{bm}
\usepackage{graphics,graphicx}

\usepackage{subfigure}
\usepackage{xcolor}

\newcommand{\be}{\begin{equation}}
\newcommand{\ee}{\end{equation}}
\newcommand{\bea}{\begin{eqnarray}}
\newcommand{\eea}{\end{eqnarray}}

\newcommand{\slsh}[1]{\not \! #1}

\begin{document}

\title{Response to an External Magnetic Field of the Decay Rate of a Neutral Scalar Field into a Charged Fermion Pair}


\author{Jorge Jaber-Urquiza$^\dagger$, Gabriella Piccinelli$^*$, Angel S\' anchez$^\dagger$}{\affiliation{
$^\dagger$Facultad de Ciencias, Universidad Nacional Aut\' onoma de M\' exico, Apartado Postal 50-542, Ciudad de M\'exico 04510, Mexico.\\
$^*$Centro Tecnol\'ogico, Facultad de Estudios Superiores Arag\' on, Universidad Nacional Aut\' onoma de M\' exico, Avenida Rancho Seco S/N, Col. Impulsora Popular Av\'icola, Nezahualc\' oyotl, Estado de M\' exico 57130, Mexico.}

\begin{abstract}
\noindent In this work we explore the effects of a weak magnetic field on the decay process of a neutral scalar boson into a pair of charged fermions in vacuum.  Since the analytical computation of the decay width needs of some approximation, following two different approaches, we study the low   and high transverse momentum limits. Our findings indicate that the magnetic field effect depends on the kinematics of the scalar particle.   

\end{abstract}


\maketitle

\section{Introduction}

Scalar fields appear in several branches of physics, in different ways, as a fundamental or composite field.
At different energy scales, they can be present from condensed matter (superconductors~\cite{LG}), compact astrophysical objects (color superconductivity~\cite{Alford}), high energy physics (Higgs physics and  heavy ion collisions~\cite{HiggsT, HiggsE, Chernodub}), up to cosmological events (where the scalar field can be a matter~\cite{Tona} and energy~\cite{yo-mera} component of the dark sector,  or the inflaton, see e.g.,~\cite{cosmoinfl}).

On the other hand, in all these processes we can also expect  the presence of a magnetic field. Thus, a natural question that emerges is: how does this magnetic field change the physics in the phenomena driven by the scalar field? Several authors have addressed this question, finding different answers and eventually discovering new phenomena such as direct and inverse Magnetic Catalysis~\cite{Miransky, imc}. Some of these works have been developed in  Inflation~\cite{warmus}; Electroweak Phase Transition and Baryogenesis~\cite{ewpt} and  QCD phase diagram in high density \cite{QCDhighmu} and temperature \cite{QCDhighT} regions  (for recent reviews, see \cite{QCDptB}).

Other answers have also come from the study of the effect of a magnetic field on the particle decay process. This effect has been studied in different situations: high intensity laser experiments~\cite{lasers}, heavy ion collisions~\cite{Chernodub,Kawaguchi,Bandyopadhyay}, compact objets~\cite{Raffelt,compactobjects} and early universe events~\cite{PS,BPS,Rudnei}.

In all these contexts, the decay processes take place in different external conditions like:
temperature, density, electromagnetic field strength and the nature of the progenitor and decay products~\cite{Borisov}. 

In the cases of study presented in the literature, two limits are typically considered: strong magnetic fields, in which case only the lowest Landau level (LLL) is taken into account, and weak magnetic fields, that allow to perform some kind of expansion series in $B$ (with $B$ the magnetic field) and keep only the lowest terms. In addition, the methods followed to calculate the decay rate and the approximations necessarily accomplished to go through the calculation vary from a work to another.

Even though  all these differences are expected to have a consequence on the process, it is nonetheless remarkable that different results can be found in the literature for the same physical process.

Besides the first works on decay processes in the presence of an external magnetic field~\cite{TsaiErber,Urrutia,Kuznetsov1,Mikheev,Erdas,Bhattacharya,Satunin}, nowadays, there is an intense activity around this subject~\cite{Bandyopadhyay,Sogut,Kawaguchi,Chistyakov,Ghosh,FelixKarbstein,Rudnei,Bali}. It is nonetheless important to further explore this area in order to discern which are the relevant physical ingredients and make the results converge when they describe the same physical situation. Moreover, to delimit the magnetic field effect on the collected data in experiments is crucial since any deviation on the expected result could be misinterpreted as new physics.

In a recent publication~\cite{PS}, we have shown that the kinematics of the progenitor particle plays an important r\^ole in the decay process. This ingredient should be taken into account since, depending on the kinematic region of interest, the decay width can be enhanced or inhibited by the external magnetic field. In the present work, we extend these ideas by exploring another  important aspect on the decay process: the spin.  

In order to explore the effect of external magnetic fields on the decay process and its possible relation with spin, 
in this work we shall study a heavy scalar boson decay into two charged fermions, in vacuum, with different approaches. In particular, we are  interested in weak field limit, in such a way that our study could be applied on particle production during late stages of peripheral heavy ion collisions~\cite{McLerran} and on the inflaton decay process in a warm inflation scenario \cite{Berera,Bastero-Gil,HM},  considering that cosmic magnetic fields observed at all scales in the universe~\cite{cosmicB,IntergalacticB}  could be primordial~\cite{SavvidyEnqv-Olesen,TurnerWidrow,PlanckB}.

The outline of this work is as follows: in Sec.\ref{sec2} we present the model and calculate the scalar self-energy in the presence of an external homogeneous magnetic field by considering the interaction of a neutral scalar with a pair of charged fermions, up to one loop; in Sec.\ref{sec3} we get the heavy boson self-energy in the weak field limit through two different approximations that take care of the kinematic along the transverse direction with respect to the magnetic field, and obtain the decay width by invoking the optical theorem; results  are presented in Sec.\ref{sec4} together with a discussion on the physical differences between these two approaches and with respect to scenarios analyzed by other authors. Finally, Sec.\ref{sec5} contains our conclusions.

\section{Model}\label{sec2}

Let us consider a model in which a neutral scalar boson $\phi$ interacts with two charged fermions $\overline{\psi}\psi$.  
The simplest form for a Lagrangian to account for this process is

\begin{equation}
\mathcal{L}=g\phi\overline{\psi}\psi.
\end{equation}

This interaction term gives rise to the Feynman diagram shown in Fig.\ref{fig1}, whose analytical expression reads
\begin{equation}
\Sigma^{^{B}}(x,y)=ig^2tr\bigg[S_F^{^B}(x,y)S_F^{^B}(y,x)\bigg].
\label{eq.conB-FeynmanRules1}
\end{equation}
To consider the effect of an external magnetic field, we use the Schwinger's proper-time method~\cite{Schwinger}, that incorporates the full magnetic contribution into the momentum dependent propagators, in such a way the propagator has the form
\begin{equation}
S_F^{^B}(x,y)=\Omega(x,y)\int \dfrac{d^4k}{(2\pi)^4}S_F^{^B}(k)e^{-ik(x-y)},
\label{eq.EcDirB-PropConf}
\end{equation}
where
\begin{equation}
S_F^{^B}(k)=\int_{0}^{\infty}\frac{ds}{\cos(eBs)} e^{-is\left(m^2-k_{\parallel}^2-k_{\perp}^2\frac{\tan(eBs)}{eBs}\right)}\left[\left(m+\slsh{k}_{\parallel}\right)e^{-ieBs\Sigma_3}+\dfrac{\slsh{k}_{\perp}}{\cos(eBs)}\right]
\label{eq.EcDirB-PropMomento}
\end{equation}
with
\begin{equation}
\Omega(x',x'')=\exp\left(-ie\int_{x''}^{x'} A_\mu(x)dx^\mu\right)
\label{eq.EcDirB-FaseSchwinger}
\end{equation}
the well-known Schwinger's phase which encodes the gauge dependence of the external magnetic field, and  $e$ and $m$ denote the charge and mass of the fermion fields  $\psi$, respectively, $B$ is the external magnetic field and $\Sigma_3=i\gamma_1\gamma_2$, with $\gamma$'s the Dirac gamma matrices. Notice that this particular form for the propagator implies that we have considered an external uniform magnetic field along $z$-direction, allowing us to  adopt  the notation $(a \cdot b)_{||} \equiv a^0 b^0 - a^3 b^3$ and $(a \cdot b)_\perp \equiv -a^1 b^1 - a^2 b^2$.

In the momentum space, the self-energy reads  
\begin{equation}
\Sigma(p)=ig^2\int\dfrac{d^4k}{(2\pi)^4}tr\left[S^B_F(k)S^B_F(k-p)\right],
\label{eq.sinB-FeynmanRules3}
\end{equation}
where the phase has canceled identically, due to a close loop of charged fermions. 
\begin{figure}[h!]
	\centering
	\includegraphics[scale=0.3]{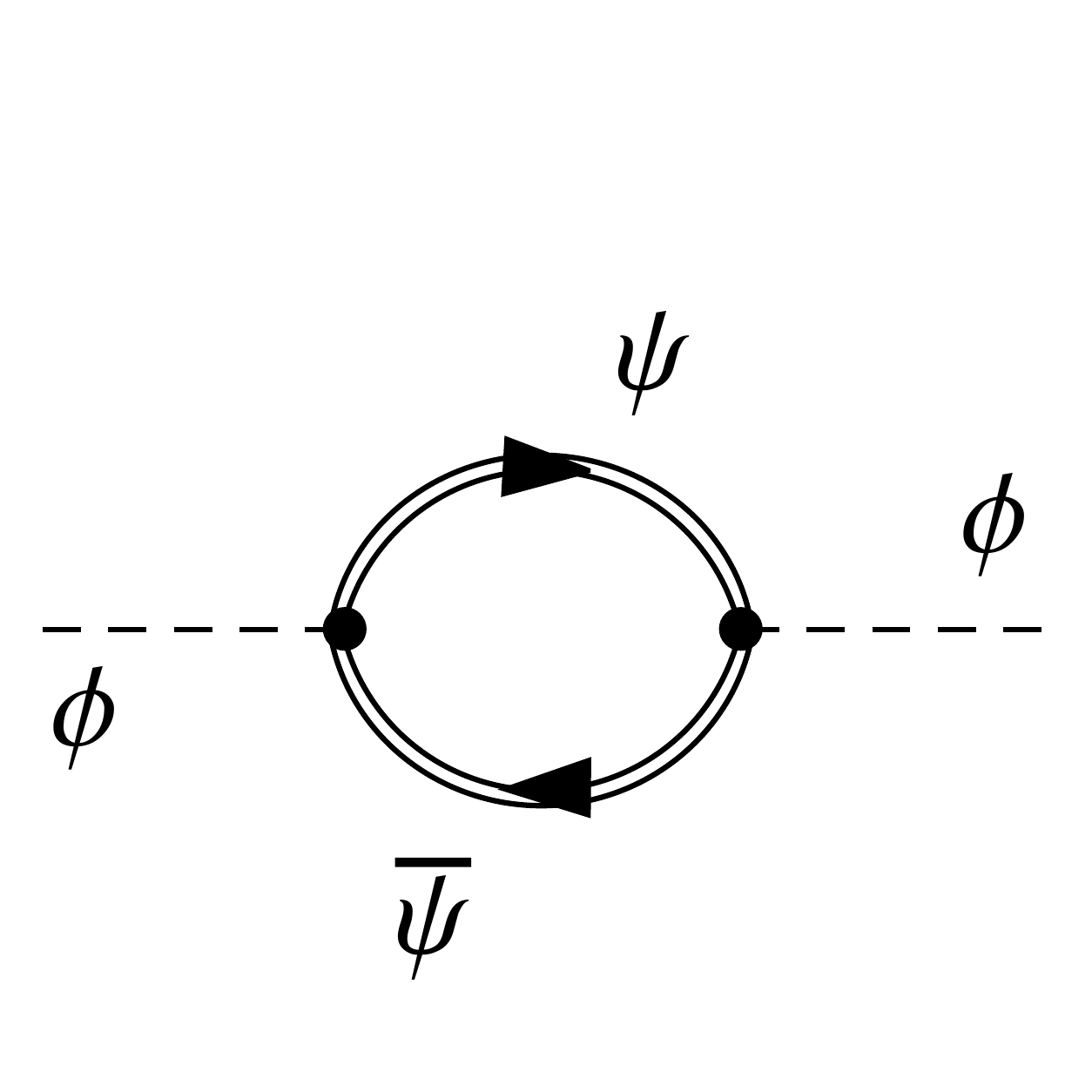}
    \caption{Leading-order contribution to the neutral scalar $\phi$  boson self-energy in a magnetic field background. $\psi's$ are charged fermions and  the double dashed lines represent their propagators dressed with the magnetic field.}
    	\label{fig1}
\end{figure}
Once we replace the propagators from Eq.~(\ref{eq.EcDirB-PropMomento}) in Eq.~(\ref{eq.sinB-FeynmanRules3}), the scalar self-energy dressed with a magnetic field reads 
\begin{equation}
\begin{split}
\Sigma^{^{B}}(p)=ig&^2\int \dfrac{d^4k}{(2\pi)^4}\int_{0}^{\infty}ds_1\int_{0}^{\infty}ds_2\dfrac{e^{-im^2(s_1+s_2)}}{\cos(eBs_1)\cos(eBs_2)} e^{is_1\left(k_{\parallel}^2+k_{\perp}^2\frac{\tan(eBs_1)}{eBs_1}\right)} e^{is_2\left((k-p)_{\parallel}^2+(k-p)_{\perp}^2\frac{\tan(eBs_2)}{eBs_2}\right)}\\
&\hspace{20mm}\times tr\bigg\{\left[\left(m+\slsh{k}_{\parallel}\right)e^{-ieBs_1\Sigma_3}+\dfrac{\slsh{k}_{\perp}}{\cos(eBs_1)}\right]\left[\left(m+\slsh{k}_{\parallel}-\slsh{p}_{\parallel}\right)e^{-ieBs_2\Sigma_3}+\dfrac{\slsh{k}_{\perp}-\slsh{p}_{\perp}}{\cos(eBs_2)}\right]\bigg\},
\end{split}
\label{eq.conB-FeynmanRules6}
\end{equation}

Now, by using the change of variables
\begin{eqnarray*}
s=s_1+s_2  \mbox{\ \ \ and \ \ \ }   v=\frac{s_1-s_2}{s_1+s_2},
\end{eqnarray*}
and performing the trace over Dirac's gamma matrices as well as the Gaussian integrals over the momenta $k$, we finally arrive at   
\begin{equation}
\begin{split}
\Sigma^{^{B}}(p)=\dfrac{g^2eB}{8\pi^2}&\int_{0}^{\infty}ds\int_{-1}^{1}dv\hspace{2mm}e^{-ism^2}e^{i\frac{s(1-v^2)}{4}{p_{\parallel}}^2}e^{\frac{i}{eB}\frac{\cos(eBsv)-\cos(eBs)}{2\sin(eBs)}{p_\perp}^2}\\
&\hspace{5mm}\times\bigg[\dfrac{1}{\tan(eBs)}\left(m^2+\dfrac{i}{s}-\dfrac{1-v^2}{4}{p_{\parallel}}^2\right)+\dfrac{ieB}{\sin^2(eBs)}-\dfrac{\cos(eBsv)-\cos(eBs)}{2\sin^3(eBs)}{p_{\perp}}^2\bigg],
\end{split}
\label{eq.conB-AutoE.4}
\end{equation}
which accounts for the effect of a magnetic field  on the scalar self-energy due to a loop of charged fermions. 
Eq.~(\ref{eq.conB-AutoE.4}) is an exact result, since we have not made any approximation on the magnetic field strength. However, the remaining integrals can not be calculated analytically because of the involved form of the self-energy. To gain some insight about the magnetic field effect on the analytical structure of the self-energy, in what follows we shall explore two approximations.

\section{Magnetic Field Effect on Scalar decay into Fermions}\label{sec3}

Taking into account that in Eq.~(\ref{eq.conB-AutoE.4}) the physical scales are the progenitor particle momentum $p$, the daughter particle mass $m$ and the magnetic strength interaction $eB$, the approximations that can be done depend on the hierarchy of scales among these quantities. In particular, as we mentioned in the introduction, we are interested in weak magnetic fields with respect to the mass $m$, {\it i.e}., $eB \ll m^2$. In this regime, we still have two possibilities depending on the third physical scale involved in the self-energy, the transverse momentum. As it can be seen in Eq.~(\ref{eq.conB-AutoE.4}), there is an exponential factor that involves the combination of the transverse momentum and the magnetic field. In this way, we should be careful with the expansion of this term.  In what follows, we study two different approximations depending on the progenitor particle kinematics.
 
\subsection{Weak magnetic field and low transverse momentum limit}

Let us start by considering the limit in which both the magnetic field and the decaying particle transverse momentum are low and of the same order of magnitude. In this case, all the factors in the integrand of Eq.~(\ref{eq.conB-AutoE.4}) can be expanded up to $(eB)^2$ terms and the self-energy can be written as

\begin{equation}
\Sigma^{^{B}}(p)\simeq\Sigma(0)+\tilde{\Sigma}(p)+\tilde{\Sigma}^{^B}(p),
\label{eq.ConB-AutoB}
\end{equation}

\noindent where the first two terms correspond to the self-energy in vacuum and are given by\

\begin{equation}
\Sigma(0)=\dfrac{g^2}{4\pi^2}\int_{0}^{\infty}\dfrac{ds}{s}\hspace{1mm}e^{-im^2s}\left[\dfrac{2i}{s}+m^2\right],
\label{selfvac}
\end{equation}

\vspace{4mm}

\noindent which is the contribution that contains the divergencies associated with the mass $m$ and the coupling constant $g$, and  \\

\begin{equation}
\tilde{\Sigma}(p)=\dfrac{g^2}{8\pi^2}\bigg[\bigg(\dfrac{p^4}{2}-2m^2p^2\bigg)\int_{-1}^{1}dv\dfrac{v^2}{(1-v^2)p^2-4m^2}+\dfrac{p^4}{2}\int_{-1}^{1}dv\dfrac{v^4}{(1-v^2)p^2-4m^2}-p^2\int_{0}^{\infty}\dfrac{ds}{s}e^{-ism^2}\bigg],
\label{selffinite}
\end{equation}
that is composed by two finite terms (the first two terms) and a  divergent term (the last one) related to the wave function. Since these divergencies are related with physical scales, all of them are reals and will not affect the imaginary part in which we are interested here.

\vspace{4mm}

\noindent{Finally, the magnetic contribution in Eq.~(\ref{eq.ConB-AutoB}) is\\

\begin{equation}
\begin{split}
\tilde\Sigma^{^{B}}(p)=\dfrac{g^2}{8\pi^2}\bigg[&\dfrac{16}{3}(eB)^2m^2\int_{-1}^{1}\dfrac{dv}{((1-v^2)p^2-4m^2)^2}-\dfrac{4}{3}(eB)^2{p_\parallel}^2\int_{-1}^{1}dv\dfrac{1-v^2}{((1-v^2)p^2-4m^2)^2}\\
&\hspace{5mm}+\dfrac{8}{3}(eB)^2{p_\perp}^2m^2\int_{-1}^{1}dv\dfrac{(1-v^2)^2}{((1-v^2)p^2-4m^2)^3}-\dfrac{2}{3}(eB)^2p^2{p_\perp}^2\int_{-1}^{1}dv\dfrac{(1-v^2)^3}{((1-v^2)p^2-4m^2)^3}\\
&\hspace{80mm}+\dfrac{1}{3}(eB)^2{p_\perp}^2\int_{-1}^{1}dv\dfrac{(1-v^2)(7-3v)}{((1-v^2)p^2-4m^2)^2}\bigg].
\end{split}
\label{eq.ConB-AutoBR}
\end{equation}
Note that by doing this expansion we have taken into consideration that the main contribution to the integral comes from the region $eBs \ll 1$~\cite{TsaiErber,Urrutia}. In Eqs.~(\ref{selfvac}, \ref{selffinite}, \ref{eq.ConB-AutoBR}), the integrals over $v$ or $s$ have been performed when the operation became trivial.

In order to calculate the decaying rate, we invoke the optical theorem that relates the imaginary part of the self-energy and the decay rate as
\cite{peskin}
\begin{equation}
\Gamma=-\dfrac{\Im\left(\Sigma(p)\right)}{\sqrt{{\vec{p}}^2+M^2}}\hspace{1mm}.
\label{eq.Pre.decayrate}
\end{equation}
where the imaginary part of the self-energy up to one loop is represented by cutting the Feynman diagram, as shown in Fig.~\ref{fig.AutoE2}.
\begin{figure}[h!]
	\centering
	\includegraphics[scale=0.3]{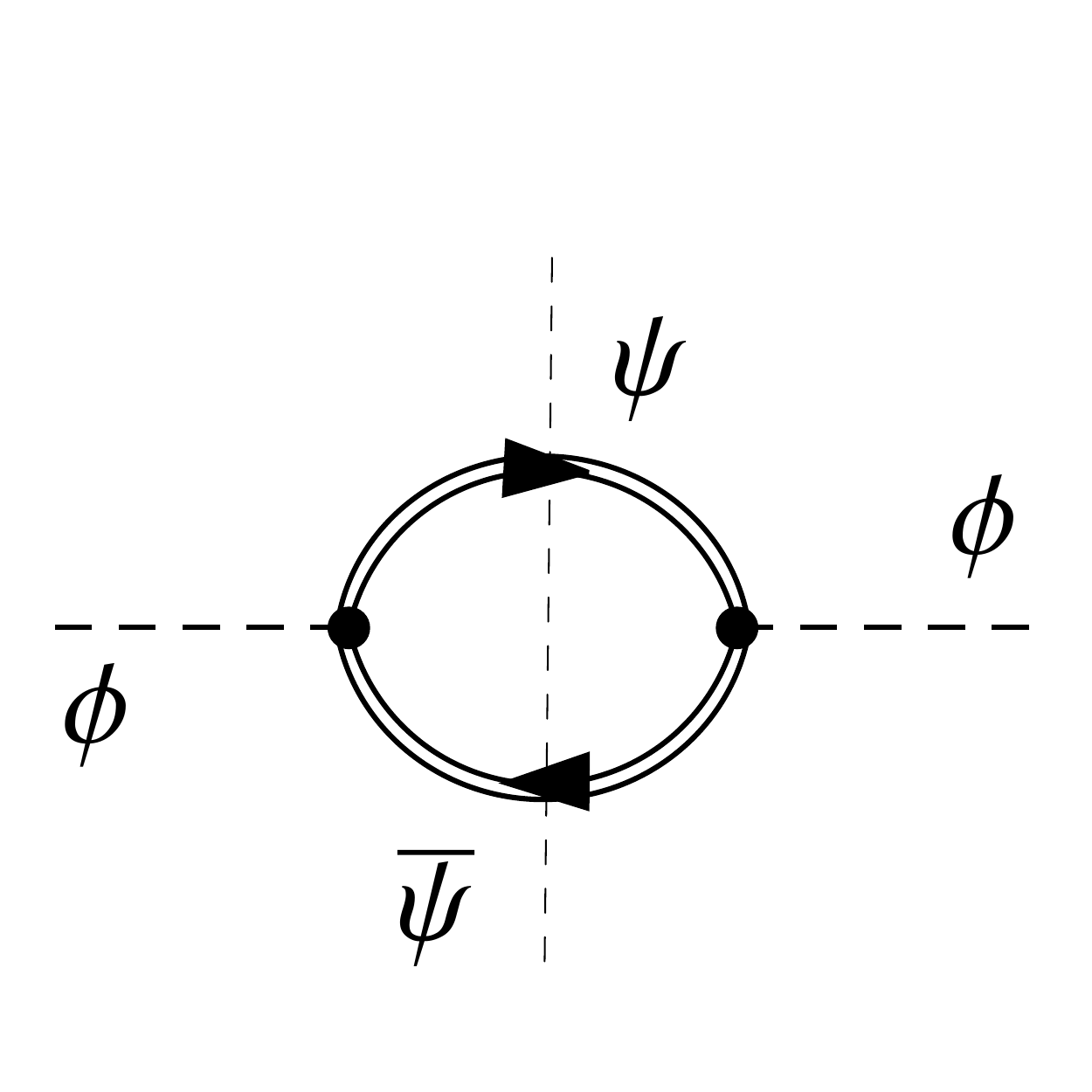}
	\caption{Schematic representation of the imaginary part of the self-energy, obtained by cutting the one-loop Feynman diagram.}
    	\label{fig.AutoE2}
\end{figure}

Replacing  Eq.~(\ref{eq.ConB-AutoB}) into Eq.~(\ref{eq.Pre.decayrate}) and bearing in mind that the mass $m$ has an $i\epsilon$ term that warranties the causality in the propagator in Eq.~(\ref{eq.EcDirB-PropMomento}), the calculation of the imaginary part can be easily done by using the identities~\cite{butkov,stone} 
\begin{equation}
\lim_{\epsilon \to 0}\Im\left(\dfrac{1}{(x+i\epsilon)^n}\right)=\dfrac{(-1)^{n+1}}{n!}\dfrac{d^n \delta(x)}{dx^n},
\label{eq.ConB-DeltaDiracDerivadas0}
\end{equation}
and\\
\begin{equation}
\int_{.\infty}^{\infty}\dfrac{d^m\delta(x-a)}{dx^m}f(x)dx=(-1)^m\dfrac{d^mf(x)}{dx^m}\bigg|_{x=a}\hspace{1mm}.
\label{eq.ConB-DeltaDiracDerivadas}
\end{equation}
In such a way, we obtain for the decay rate \\
\begin{equation}
\begin{split}
\Gamma^B=\dfrac{2g^2}{16\pi}\bigg[&\dfrac{(p^2-4m^2)^{3/2}}{p}-\dfrac{24(eB)^2m^4{p_\perp}^2}{p^5(p^2-4m^2)^{3/2}}-\dfrac{16(eB)^2m^2{p_\perp}^2}{3p^5}\dfrac{12m^4-7m^2p^2+p^4}{(p^2-4m^2)^{5/2}}\\
&\hspace{50mm}+\dfrac{4(eB)^2}{3p^3}\dfrac{4m^2p^2+2m^2{p_\perp}^2+2p^2{p_\perp}^2-p^4}{(p^2-4m^2)^{3/2}}\bigg]\dfrac{\Theta(p^2-4m^2)}{\sqrt{{\vec{p}}^2+M^2}}\hspace{1mm},
\end{split}
\label{eq.conB-TasaAproxFinal}
\end{equation}

In particular, for an on-shell scalar particle in its rest frame ($p^2=M^2$), the decay width reads
\begin{equation}
\Gamma^B=\dfrac{2g^2}{16\pi}\bigg[\dfrac{(M^2-4m^2)^{3/2}}{M^2}-\dfrac{4(eB)^2}{3M^2\sqrt{M^2-4m^2}}\bigg]\Theta(M^2-4m^2).
\label{eq.conB-TasaAproxFinal2}
\end{equation}

The apparent divergencies near the threshold $p=2m$ in the second term of this equation, as well as in several terms of Eq.~(\ref{eq.conB-TasaAproxFinal}), are in fact not present since, in order to be consistent with the weak field limit expansion, it is required that the magnetic field strength goes to zero faster than the terms that contain the threshold information. This statement comes from the fact that in the expansion shown in Eq.(\ref{eq.conB-TasaAproxFinal}), the expansion parameter  is $eB/(p^2-4m^2)$.  This follows from the fact  that in Eq.~(\ref{eq.conB-AutoE.4}), in the absence of a magnetic field, we have two  different physical scales: $m^2$ and $(p^2-4m^2)$, which correspond to the extreme values in the argument of the exponential. The last one is the relevant scale for the development of the self-energy  imaginary part (see also \cite{Rudnei}).

\subsection{Weak magnetic field and high transverse momentum limit}

Let us now explore the other weak field limit: $eB\ll m^2 \ll p_\perp^2$. In this case we must carry out a Taylor expansion with care~\cite{TsaiErber}: the argument in the exponential that involves the magnetic field  is expanded up to $(eBs)^2$, however, since there are terms that contain a factor $p_\perp^2$ that can be large (known as crossed field approximation), then, the exponential itself cannot be expanded in powers of $eBs$. This argument does not apply to the coefficient in front of the exponential since the transverse momentum is weighted by the total momentum, thus, the leading contribution is of the order $\mathcal{O}(1)$. Bearing this in mind, we get
\begin{equation}
\begin{split}
\Sigma^{^{B}}(p)\simeq\dfrac{g^2}{8\pi^2}&\int_{0}^{\infty}ds\int_{-1}^{1}dv\hspace{2mm}e^{-ism^2}e^{i\frac{s(1-v^2)}{4}p^2}e^{is^3\frac{(eB)^2}{48}(1-v^2)^2{p_{\perp}}^2}\\
&\hspace{3mm}\times\bigg[\dfrac{m^2}{s}+\dfrac{2i}{s^2}-\dfrac{1-v^2}{4s}p^2-\dfrac{m^2(eB)^2s}{3}+\dfrac{(eB)^2s}{12}(1-v^2){p_{\parallel}}^2-\dfrac{(eB)^2s}{48}(1-v^2)(5-v^2){p_{\perp}}^2\bigg].
\end{split}
\label{eq.conB-AutoE.Tsai.2}
\end{equation}

In a similar way as we proceeded in the previous section, we can decompose the self-energy as 
\begin{equation}
\Sigma^{^{B}}(p)\simeq\Sigma(0)+\tilde{\Sigma}^{^B}(p),
\label{eq.ConB-AutoB.2}
\end{equation}
with $\Sigma(0)$  the same term as in Eq.~(\ref{selfvac}), and 
\begin{equation}
\begin{split}
\tilde{\Sigma}^{^B}(p)=\dfrac{g^2}{8\pi^2}\bigg[&-p^2\int_{0}^{\infty}\dfrac{ds}{s}e^{-ism^2}-\dfrac{(eB)^2}{3}\left(m^2-\dfrac{{p_\parallel}^2}{4}+\dfrac{5{p_\perp}^2}{16}\right)V_0S_1\\
&+\dfrac{i}{2}\left(m^2p^2-\dfrac{p^4}{4}\right)V_2S_0-\dfrac{(eB)^2}{12}\left({p_\parallel}^2+\dfrac{{p_\perp}^2}{2}\right)V_2S_1\\
&+i\dfrac{(eB)^2}{12}{p_\perp}^2\left(m^2-\dfrac{p^2}{4}\right)V_2S_2-i\dfrac{p^4}{8}V_4S_0\\
&+\dfrac{7(eB)^2}{48}{p_\perp}^2V_4S_1-i\dfrac{m^2(eB)^2}{12}{p_\perp}^2V_4S_2+i\dfrac{(eB)^2}{48}p^2{p_\perp}^2V_6S_2\bigg],
\end{split}
\label{eq.conB-AutoE(p).Tsai.1}
\end{equation}
where, for simplicity, we have introduced the notation
\begin{equation}
V_nS_m\equiv\int_{-1}^{1}dv\hspace{1mm}v^n\int_{0}^{\infty}ds\hspace{1mm}s^me^{is\left(\frac{1-v^2}{4}p^2-m^2+s^2\frac{(eB)^2}{48}(1-v^2)^2{p_\perp}^2\right)}.
\label{eq.ConB-Def.V_nS_m}
\end{equation}
It is worth to notice that in this case it was not possible to identify the $\tilde{\Sigma}(p)$ term due to the fact that the integrals include a new exponential factor. 

Once we replace  Eq.~(\ref{eq.ConB-AutoB.2}) in Eq.~(\ref{eq.Pre.decayrate}), we arrive at
\begin{equation}
\begin{split}
\Gamma^{^B}=\dfrac{g^2}{16\pi}\dfrac{8\sqrt[3]{2}}{(eB)^{2/3}{p_\perp}^{2/3}}\bigg[&-\dfrac{p^2}{2}\int_{_{\sqrt{1-\frac{4m^2}{p^2}}}}^{1}dv(1-v^2)^{-2/3}v^2\left(\frac{1+v^2}{4}p^2-m^2\right)Ai(x)\\
&+\dfrac{4}{3}\int_{_{\sqrt{1-\frac{4m^2}{p^2}}}}^{1}dv(1-v^2)^{-5/3}v^2\left(\frac{1+v^2}{4}p^2-m^2\right)\left(\frac{1-v^2}{4}p^2-m^2\right)Ai(x)\\
&+\dfrac{2\sqrt[3]{2}(eB)^{4/3}}{3{p_\perp}^{2/3}}\int_{_{\sqrt{1-\frac{4m^2}{p^2}}}}^{1}dv(1-v^2)^{-4/3}\left(\frac{1-v^2}{4}{p_\parallel}^2-\frac{(1-v^2)(7v^2-5)}{16}{p_\perp}^2-m^2\right)Ai'(x)\bigg]\\
&\hspace{115mm}\times\dfrac{\Theta(p^2-4m^2)}{\sqrt{{\vec{p}}^2+M^2}},
\end{split}
\label{eq.conB-TasaTsaiFinal}
\end{equation}
where the terms that involved the integration over $s$ have been identified with the integral representation of the first kind Airy functions\\
\begin{equation}
Ai(x)=\dfrac{1}{\pi}\int_{0}^{\infty}\cos\left(xt+\frac{t^3}{3}\right)dt
\label{eq.ConB-Airy}
\end{equation}
 and its derivatives~\cite{ovalle}, 
 with 
\begin{equation}
  x=\frac{m^2-\frac{1-v^2}{4}p^2}{\left(\frac{-p_\perp^2(eB)^2(1-v^2)^2}{16}\right)^{1/3} }.
\end{equation}

Note that the zero magnetic field limit can be recovered by using the Dirac delta representation~\cite{ovalle}
\begin{equation*}
\delta(x)\equiv\lim_{\varepsilon \to 0}\frac{1}{\varepsilon}Ai\left(\frac{x}{\varepsilon}\right)
\mbox{ \ \ \ with \ \ \ }
\varepsilon\equiv\left(\dfrac{(eB)^2}{m^4}\right)^{\frac{1}{3}}.
\label{eq.ConB-DefDeltaAiry}
\end{equation*}

It is worth to keep in mind that this approximation does not allow us to study the decay process in the particle rest frame since, by construction,  $p_\perp$ is large. 

In the next section we present our results and discuss the r\^ole played by the decaying particle kinematics: $p_\perp\sim m$ and $p_\perp\gg m$.

\section{Results and Discussion}\label{sec4}

The decay rate as a function of the magnetic field normalized with the fermion mass $m$, for different transverse momentum values, in the low momentum approximation, given by Eq.~(\ref{eq.conB-TasaAproxFinal}), is shown In Fig.~\ref{fig.DecayRateVsEB-Week}. In this figure and the following ones we ignore the factor $g^2/16 \pi$. We can notice that the magnetic field slightly suppresses the decay process. The effect of the dilatation of time, through the Lorentz factor, can also be observed, and becomes evident at zero magnetic field. It is worth to notice that each curve represents the decay rate measured by observers in different reference frames.  
\begin{figure}[h!]
	\centering
	\includegraphics[scale=0.85]{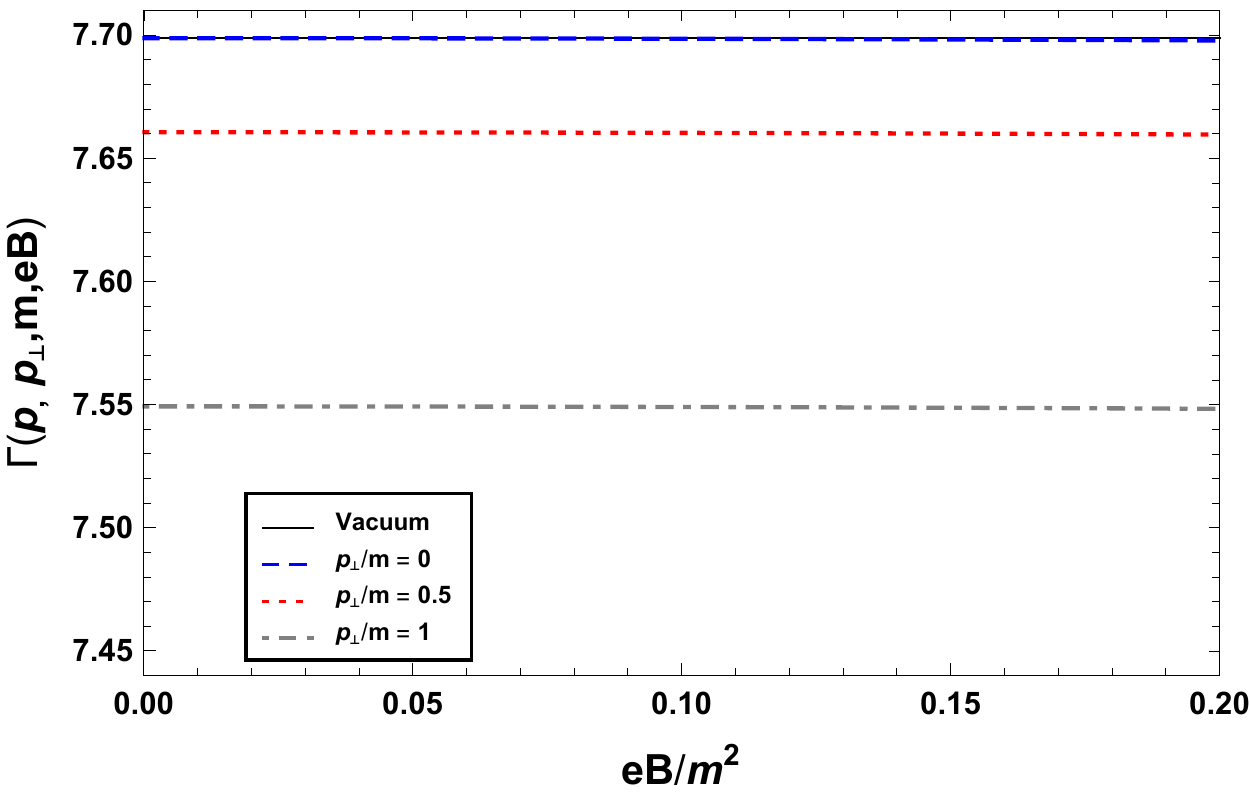}
	\caption{Decay rate (in units of energy), as a function of the magnetic field strength, for different values of the transverse momentum, for the low momentum approximation, taking $p/m=5$.}
	\label{fig.DecayRateVsEB-Week}
\end{figure}

In order to quantify the magnetic field effect on the decay process it is convenient to compare it with the decay rate in vacuum,  
\begin{equation}
\Delta\Gamma(p,p_\perp,m,eB)\equiv\dfrac{\Gamma^{^B}(p,p_\perp,m,eB)}{\Gamma_{\text{\textit{vac}}}(p,p_\perp,m)}\hspace{1mm},
\label{eq.compara-DeltaGamma}
\end{equation}
where $\Gamma_{\text{\textit{vac}}}(p,p_\perp,m)$ corresponds to the first term in Eq.~(\ref{eq.conB-TasaAproxFinal}). For definiteness, we dub this expression as the decay response to the external magnetic field.   In this way we also eliminate the relativistic effect due to dilatation of time and the dependence on $eB$ can more easily be seen. In Fig.~\ref{fig.DeltaDecayRateVsEB-Week}, we show the behavior of $\Delta \Gamma$ as a function of $eB$, for different values of the transverse momentum.
\begin{figure}[h!]
	\centering
	\includegraphics[scale=0.85]{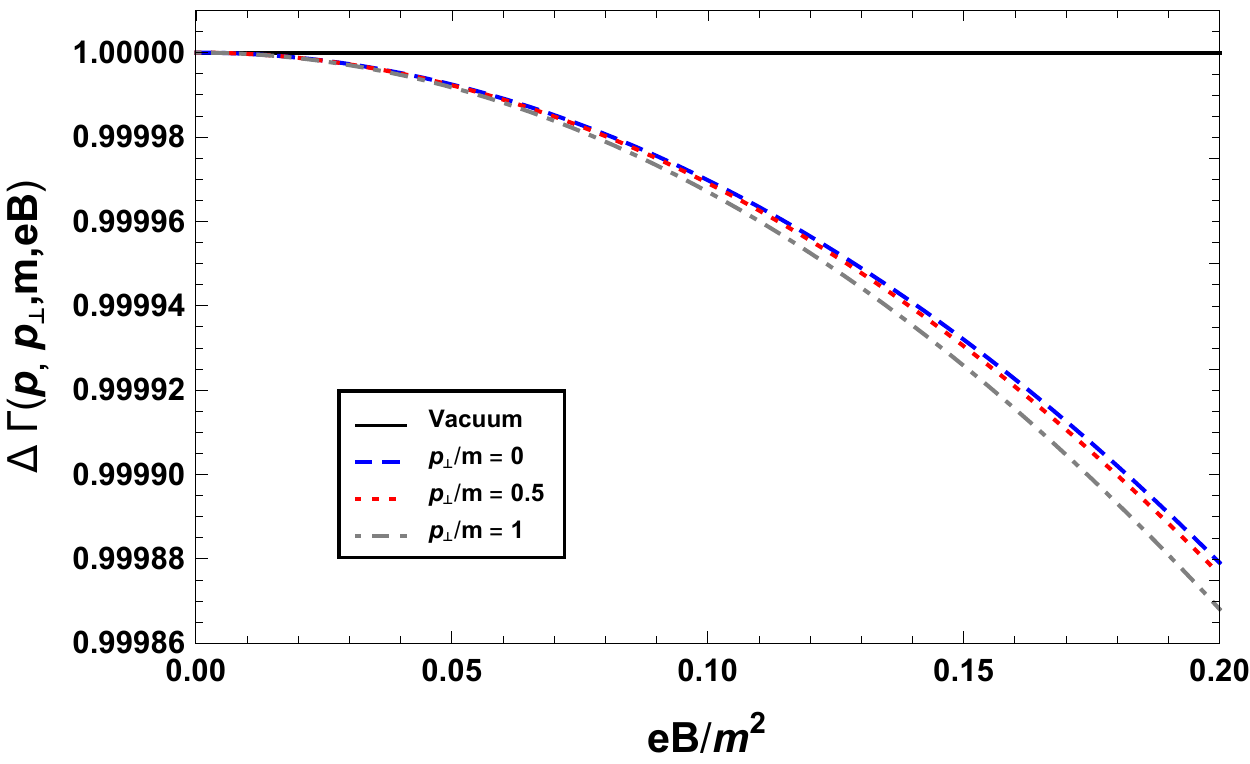}
	\caption{Decay response to the external magnetic field as a function of the magnetic field strength, for different values of the transverse momentum, for the low momentum approximation, taking $p/m=5$.}
	\label{fig.DeltaDecayRateVsEB-Week}
\end{figure}

\begin{figure}[h!]
	\centering
	\includegraphics[scale=0.85]{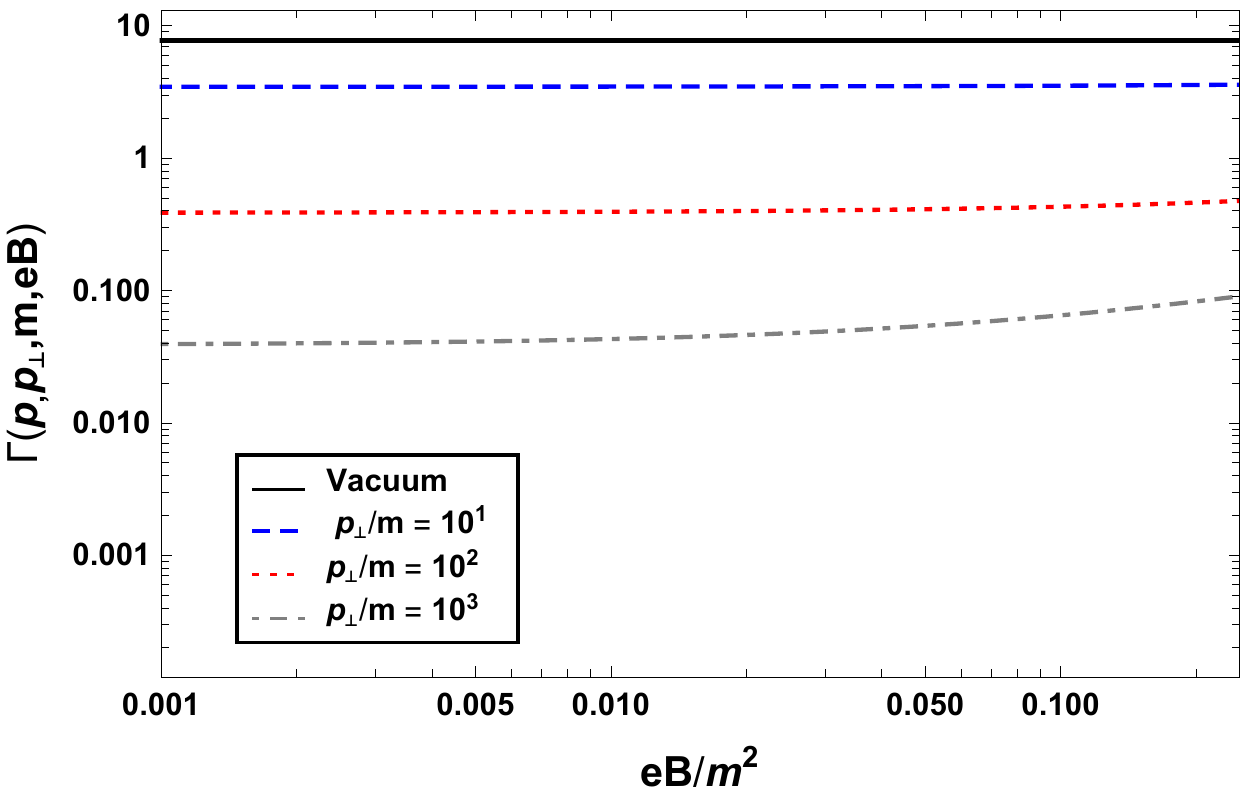}
	\caption{Decay width (in units of energy) as a function of the magnetic field for different values of the transverse momentum, for the high momentum approximation, taking $p/m=5$.}
	\label{fig.DecayRateVsEB-Tsai}
\end{figure}

\begin{figure}[h!]
	\centering
	\includegraphics[scale=0.85]{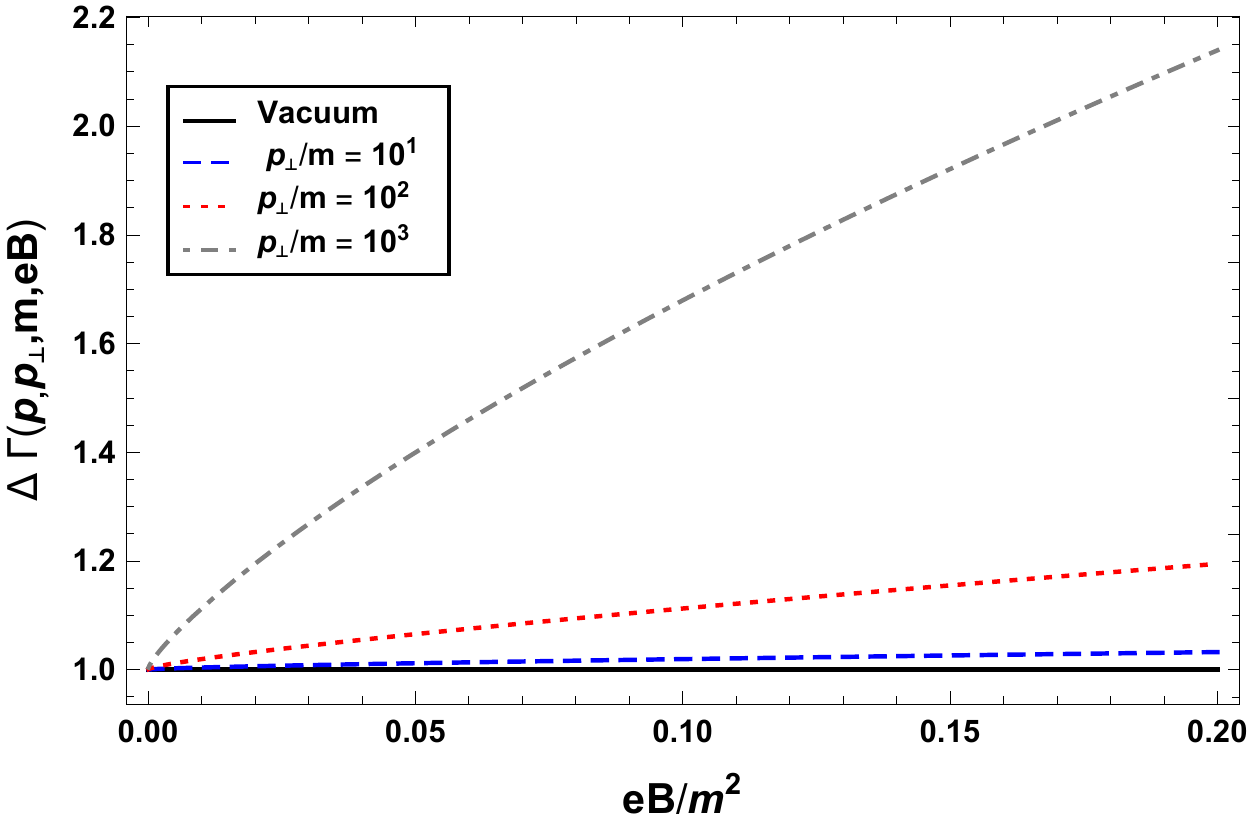}
	\caption{Decay response to the external magnetic field as a function of the magnetic field strength, for different values of the transverse momentum, in the high momentum approximation, taking $p/m=5$.}
	\label{fig.DeltaDecayRateVsEB-Tsai}
\end{figure}
The behavior of the decay rate with the magnetic field, for the high momentum approximation, can be seen in Fig.~\ref{fig.DecayRateVsEB-Tsai}. We plotted a log-log graph, in order to remark the variation of the decay rate, which is highly affected by the Lorentz factor in this regime.
Here, we can notice that the decay process follows a different behavior: the increase of the magnetic field enhances the  decay width and this is magnified as the transverse momentum grows. From physical grounds, this behavior is expected if the progenitor particle is considered as composed of a pair of charged fermions on which  the Lorentz force acts in oposite directions, promoting the breaking of the system.
To separate the relativistic effects on the decay width and highlight the pure magnetic field contribution, in Fig.~\ref{fig.DeltaDecayRateVsEB-Tsai}, we plot $\Delta \Gamma$ as a function of $eB$ by using the high momentum approximation given by  Eq.~(\ref{eq.conB-TasaTsaiFinal}), for different values of the transverse momentum.

In order to explore the r\^ole played by the spin of the particle products in the decay process, 
in Fig.~\ref{fig.DecayRatioFermion-BosonVsEB-Tsai}, we compare the $\Delta \Gamma$ for fermions, our Eq.~(\ref{eq.compara-DeltaGamma}) for the decay width in the high momentum approximation, with the  $\Delta \Gamma$ for scalars in the same approximation, obtained in a previous work~\cite{PS}, for $p_\perp/m =10^3 $. The behavior in Fig.~\ref{fig.DecayRatioFermion-BosonVsEB-Tsai} seems to indicate that, as the magnetic field grows, the progenitor scalar particle prefers to decay to a pair of charged fermions than to a pair of scalars, for large momenta.
More work is needed, in the different kinematical regimes, in order to define the r\^ole of the statistics of the decay products, in the presence of an external magnetic field. This work is in progress and will be published elsewhere.

\begin{figure}[h!]
	\centering
	\includegraphics[scale=0.9]{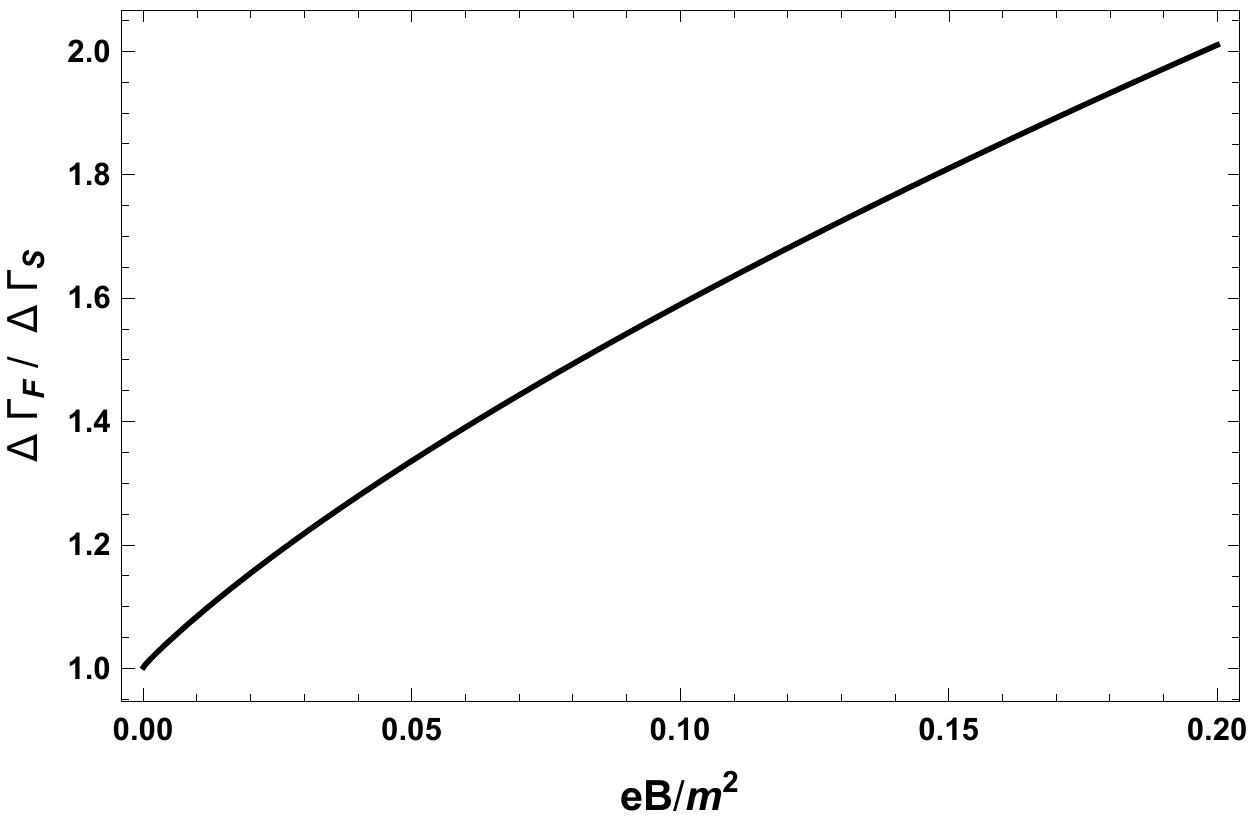}
	\caption{ Ratio of the decay responses to the external magnetic field for two different channels, $\phi\rightarrow \bar{\psi}+\psi$ and $\phi\rightarrow \varphi+\varphi$, where $\psi$ and $\varphi$  are charged  fermions and scalars, respectively. We plot this ratio as a function of the magnetic field in the high momentum approximation, for $p_\perp/m=10^3$, and, as usual, we take $p/m=5$.}
	\label{fig.DecayRatioFermion-BosonVsEB-Tsai}
\end{figure}

\section{Conclusions}\label{sec5}

In this work we have studied the magnetic field effects on the decay process  in vacuum of a neutral scalar boson into two charged fermions. Focusing on a weak magnetic field, we carried out  a perturbative approach and found that, depending on the progenitor particle kinematics, the response is inhibited or enhanced by the magnetic field. In the low transverse momentum approximation, which allows to explore the decaying particle rest frame, we observe that  the external field suppresses the decay width, meanwhile, for the high transverse momentum approximation, the effect is the opposite: as the field strength increases, the pair creation rate is enhanced.

In the low momentum approximation we found that, for a fixed transverse momentum, the decay width $\Gamma^{^B}$, as well as the response $\Delta\Gamma$, decreases as the magnetic field strength grows. This effect could be associated with the phase space reduction of the charged fermions which is caused by the increasing separation of the Landau levels with the magnetic field strength.

In the high momentum regime, we found that $\Gamma^{^B}$ increases with the magnetic field intensity, for a fixed transverse momentum. This could be explained as follows: if we imagine that the decaying scalar particle were a two charged fermion composite particle, then the Lorentz force would act in opposite directions on each forming particle, promoting the decay process. The magnetic field effect on the decay process was emphasized by considering the ratio between the full expression (containing both the vacuum and magnetic contribution) and the pure vacuum expression (Fig.~\ref{fig.DeltaDecayRateVsEB-Tsai}). In this way,  the relativistic effect related to the dilatation of time was also avoided.

As we mentioned in the introduction, many different results can be found in the literature for the effect of the magnetic field on the decay process. Since an ingredient that could be important is the spin, 
in order to shed some light on its r\^ole, we compared the response of the system to the magnetic field in two different channels:  $\phi\rightarrow \bar{\psi}+\psi$ and $\phi\rightarrow \varphi+\varphi$, where $\psi$ and $\varphi$ are charged fermions and scalars, respectively.
This was done, in the high momentum regime, by taking the ratio between the findings in the present work and the results in Ref.~\cite{PS} and shown in Fig.~\ref{fig.DecayRatioFermion-BosonVsEB-Tsai}. We observed that, in this case, the magnetic field impact is more important on the decay process to charged fermions  than to charged scalars. This results may indicate that the spin-magnetic field interaction increases the pair production in the fermion case. 

The situation studied in this work could be relevant in different physical scenarios, in particular: dilepton production in heavy-ion collisions, direct URCA processes in neutron stars, or during  an early stage of the universe evolution.

\acknowledgements
J. J. U. would like to thank to Efrain Ferrer and Vivian de la Incera  for stimulating discussions. 
Support for this work has been received in part from DGAPA-UNAM under grant numbers PAPIIT-IN117817 and PAPIIT-IA107017. 


\begin{thebibliography}{}

\bibitem{LG} V. L. Ginzburg, L. D. Landau, Zh. Eksp. Teor. Fiz.{\bf 20},  1064 (1950).

\bibitem{Alford} M. Alford, K. Rajagopal and F. Wilczek, Phys. Lett. B {\bf 422}, 247 (1998).

\bibitem{HiggsT} P. W. Higgs, Phys, Rev. Lett {\bf 13}, 508 (1964); G. S. Guralnik, C. R. Hagen and T. W. B. Kibble, Phys. rev. Lett. {\bf 13}, 585 (1964).
 
 
\bibitem{HiggsE} G. Aad. {\it et al} [ATLAS Collaboration], Phys. Lett. B. {\bf 716}, 1 (2012.); S. Chatrchyan {\it et al} [CMS Collaboration], Phys. Lett. B {\bf 716}, 30 (2012).

\bibitem{Chernodub} 
M. N., Chernodub,  Phys. Rev. D {\bf 82}, 085011 (2010).

\bibitem{Tona} 
T. Matos, F. S. Guzm\'an, L. A. Ure\~na-L\'opez, Class. Quant. Grav. {\bf 17}, 1707 (2000).

\bibitem{yo-mera} 
A. de la Macorra and G. Piccinelli,  Phys. Rev. D {\bf 61}, 123503 (2000).

\bibitem{cosmoinfl} 
A. D. Linde, {\it Particle Physics and Inflationary Cosmology}, CRC Press (1990);
V. Mukhanov, {\it Physical Foundations of Cosmology}, Cambridge University Press (2005).


\bibitem{Miransky} V. A. Miransky, {\it Dynamical Symmetry Breaking in Quantum Field Theories}, World Scientific Publishing, 1993.

\bibitem{imc}F. Preis, A. Rebhan and A. Schmitt, JHEP 1103, 033  (2011); G.S. Bali, F. Bruckmann, G. Endrodi, F. Gruber and A. Schaefer, 
JHEP 1304, 130  (2013).

\bibitem{warmus} 
G. Piccinelli, A. Sanchez, A. Ayala and A. J. Mizher,  Phys. Rev. D {\bf 90}, 83504 (2014). 

\bibitem{ewpt}
M. Giovannini and M. E. Shaposhnikov, Phys. Rev. D {\bf 57}, 2186 (1998); P. Elmfors, K. Enqvist, and K. Kainulainen, Phys. Lett. B {\bf 440}, 269 (1998); K. Kajantie, M. Laine, J. Peisa, K. Rummukainen, and M. Shaposhnikov, Nucl. Phys. B {\bf 544}, 357 (1999); A. Sanchez, A. Ayala and G. Piccinelli, Phys. Rev. D {\bf 75}, 043004 (2007); A. Ayala, G. Piccinelli, G. Pallares. Phys. Rev. D {\bf 66}, 103503 (2002).

\bibitem{QCDhighmu}
E. J. Ferrer, V. de la Incera and C. Manuel,  Phys. Rev. Lett. {\bf 95}, 152002 (2005).
 
 \bibitem{QCDhighT}
A. Ayala {\it et al}, Phys. Rev. D {\bf 92}, 096011 (2015); A. Ayala, S. Hern\'andez-Ortiz and L. A. Hern\'andez, Rev. Mex. Fis. {\bf 64} 302 (2018)

\bibitem{QCDptB}
V. A. Miransky and I. A. Shovkovy,  Phys. Rep. {\bf 576}, 1 (2015); J. O. Andersen, W. R. Naylor and A. Tranberg, Rev. Mod. Phys. {\bf 88}, 025001 (2016).

\bibitem{lasers}
I. Akal, S. Villalba-Ch\'avez and C. M\"uller, Phys. Rev. D {\bf 90}, 113004 (2014); 
A. Gonoskov, I. Gonoskov, C. Harvey, A. Ilderton, A. Kim, M. Marklund and A. Sergeev,  Phys. Rev. Lett. {\bf 111}, 60404 (2013); 
S. Meuren, K. Z. Hatsagortsyan, C. H. Keitel and A. Di Piazza, Phys. Rev. D {\bf 91}, 13009 (2015). 




\bibitem{Kawaguchi} 
M. Kawaguchi and S. Matsuzaki. Eur. Phys. J. A {\bf 53}, 12254-1 (2017).

\bibitem{Bandyopadhyay}
A. Bandyopadhyay and S. Mallik, Eur. Phys. J. C {\bf 77}, 771 (2017).


\bibitem{Raffelt} 
G. G. Raffelt, {\it Stars as Laboratories for Fundamental Physics}, (University of Chicago Press, Chicago, 1996).

\bibitem{compactobjects}
A. K. Harding, D. Lai, Rep. Prog. Phys. {\bf 69}, 2631 (2006);
J. M. Lattimer, M. Prakash, Phys. Rep. {\bf 442}, 109 (2007);
A. Y. Potekhin, Phys. Usp. {\bf 53}, 1235 (2010), Usp. Fiz. Nauk {\bf 180}, 1279 (2010);
D. Lai, Space Sci. Rev. {\bf 191}, 13 (2015).

\bibitem{PS}
G. Piccinelli and A. Sanchez, Phys.Rev. D {\bf 96}, 076014 (2017).

\bibitem{BPS}
M. Bastero-Gil, G. Piccinelli and A. Sanchez,  Astron. Nachr. {\bf 336}, 805 (2015). 


\bibitem{Rudnei}
A. Bandyopadhyay, R. L.S. Farias, R. O. Ramos, e-print: arXiv:1807.06515 


\bibitem{Bali}
G. S. Bali, B. B. Brandt, G. Endr\H{o}di and B. Gl{\"a}\ss le, e-print: arXiv:1710.01502.

\bibitem{Borisov} 
A. V. Borisov, A. S. Vshivtsev, V. C. Zhukovskii and P. A. Eminov,  Usp. Fiz. Nauk  {\bf 167}, 241 (1997).


\bibitem{TsaiErber} 
W. Tsai and T. Erber,  Phys. Rev. D {\bf 12}, 1132  (1975); W. Tsai and T. Erber,  Phys. Rev. D {\bf 10}, 492 (1974).

\bibitem{Urrutia}
L. Urrutia, Phys. Rev. D {\bf 17}, 1977  (1978).

\bibitem{Kuznetsov1}
A. V. Kuznetsov, N. V. Mikheev and L. A. Vassilevskaya, Phys. Lett. B {\bf 427}, 105 (1998);
A. V. Kuznetsov, N. V. Mikheev, G. G. Raffelt and L. A. Vassilevskaya, Phys. Rev. D{\bf 73}, 023001 (2006). 

\bibitem{Mikheev}  
N. V. Mikheev and  N. V. Chistyakov, JETP Lett. {\bf 73}, 642 (2001).
 
\bibitem{Erdas} 
A. Erdas and M. Lissia,  Phys. Rev. D {\bf 67}, 033001 (2003).         

\bibitem{Bhattacharya} 
K. Bhattacharya and S. Sahu, Eur. Phys. J. C {\bf 62}, 481  (2009).  

\bibitem{Satunin} 
P. Satunin,  Phys. Rev. D {\bf 87}, 105015 (2013).
       
 \bibitem{FelixKarbstein}
F.  Karbstein, Phys. Rev. D {\bf 88}, 085033 (2013).       
       
\bibitem{Sogut} 
K. Sogut, H. Yanar and A. Havare.  e-Print: arXiv:1703.07776 [hep-th]. 
         
\bibitem{Chistyakov}
M. V. Chistyakov,  A. V. Kutznetzov and N. V. Mikheev, Phys. Lett. B {\bf 434}, 67 (1998).

\bibitem{Ghosh}
S. Ghosh, A. Mukherjee,  M. Mandal, S. Sarkar  and P. Roy, e-Print: arXiv:1704.05319 [hep-ph]. 


\bibitem{McLerran}
D. E. Kharzeev, L. D. McLerran and H. J. Warringa, Nuclear Physics A 803, 227 (2008);
V. V. Skokov, A. Yu. Illarionov and V. D. Toneev, Int. J. Mod. Phys. A 24 5925 (2009).

\bibitem{Berera} 
A. Berera and L. Z. Fang, Phys. Rev. Lett. {\bf 74}, 1912  (1995); A. Berera,  Phys. Rev. Lett. {\bf 75}, 3218 (1995); A. Berera and R. O. Ramos, Phys. Rev. D {\bf 63}, 103509 (2001).

\bibitem{Bastero-Gil} 
M. Bastero-Gil, A. Berera and R. O. Ramos,  JCAP {\bf 09}, 033 (2011).  

\bibitem{HM} 
L. M. Hall and I. G. Moss, Phys. Rev. D {\bf 71}, 023514 (2005).

\bibitem{cosmicB} 
T. E. Clarke, P. P. Kronberg and H. Boehringer,  Astrophys. J. {\bf 547}, L111 (2001); 
K. Dolag, M. Kachelriess, S. Ostapchenko and R. Tomas, Astrophys. J. Lett. {\bf 727}, L4 (2011); 
R. Durrer and  A. Neronov,  Astron. Astrophys. Rev. {\bf 21}, 62 (2013); 
E. J. Kim, P. P. Kronberg,  G. Giovannini and T. Venturi, Nature {\bf 341}, 720 (1989); 
A. Bonafede, L. Feretti, M. Murgia, F. Govoni, G. Giovannini, D. Dallacasa, K. Dolag and  G.B. Taylor, A\&A  {\bf 513},  A30  (2010); 
Y. Xu, P. P. Kronberg, S. Habib and  Q. W. Dufton, Astrophys. J. {\bf 637}, 19 (2006).

\bibitem{IntergalacticB} 
A. Neronov and A. Vovk, Science {\bf 328}, 73 (2010); 
F. Tavecchio, G. Ghisellini, L. Foschini, G. Bonnoli, G. Ghirlanda and P. Coppi, Mon. Not. R. Astron. Soc. {\bf 406}, L70 (2010).
 
\bibitem{SavvidyEnqv-Olesen}
S. Matinyan and G. Savvidy, Nucl. Phys. B {\bf 134}, 539 (1978); 
K. Enqvist and P. Olesen, Phys. Lett. B {\bf 329}, 195 (1994).

\bibitem{TurnerWidrow}
M. S. Turner and L. M. Widrow, Phys. Rev. D {\bf 37}, 2743 (1988).

\bibitem{PlanckB} 
P. A. R. Ade, {\it et al.} A\&A {\bf 594}, A19 (2016).

\bibitem{Schwinger} J. Schwinger,  Phys. Rev. {\bf 82}, 664 (1951).

\bibitem{peskin}
M. E. Peskin,  {\it An Introduction to Quantum Field Theory}, (Westview Press, 1995); 
V.  B. Berestetskii, L. P. Pitaevskii and E. M. Lifshitz, {\it Quantum Electrodynamics}, (Pergamon Press, 1982).

  \bibitem{butkov} 
E. Butkov, {\it Mathematical Physics}, (Addison-Wesley, 1968).

\bibitem{stone} 
M. Stone and P. Goldbart, {\it Mathematics for Physics: A Guided Tour for Graduate Students}, (Cambridge University Press, 2009).

\bibitem{ovalle}
O. Vall\'ee and M. Soares, {\it Airy Functions and Applications to Physics}, (Imperial College Press, 2004);
I. S. Gradshteyn and I. M. Ryzhik, {\it Table of Integrals, Series, and Products}, edited by A. Jeffrey and D. Zwillinger (Academic Press, 2013).
  
  



  
\end{thebibliography}
\end{document}